\documentclass[preprint,12pt]{elsarticle}

\usepackage{color}
\usepackage{amssymb}
\usepackage{epsfig}
\usepackage{epstopdf}
\usepackage{graphicx,color}
\usepackage[usenames,dvipsnames]{xcolor}
\usepackage[section]{placeins}
\usepackage{amsmath}
\usepackage{subfigure}
\usepackage[normalem]{ulem}

\usepackage{soul}

\usepackage{mathtools}
\usepackage{mathrsfs}
\usepackage{bm}
\usepackage{relsize}
\usepackage{indentfirst}

\newcommand{\cd}{\makebox[0.08cm]{$\cdot$}}

\journal{Physics Letters B}

\begin{document}
\begin{frontmatter}
\title{Bound state equation for the Nakanishi weight function}
\author{J.~Carbonell}
\address{Institut de Physique NucleÌaire, Universit\'e Paris-Sud, IN2P3-CNRS, 91406 Orsay Cedex, France}
\author{T.~Frederico}
\address{Instituto Tecnol\'ogico de Aeron\'autica, DCTA,
12228-900, S. Jos\'e dos Campos,~Brazil.}
\author{V.A.~Karmanov}
\address{Lebedev Physical Institute, Leninsky Prospekt 53, 119991 Moscow, Russia}
\date{\today}
\begin{abstract}
The bound state Bethe-Salpeter amplitude was expressed  by Nakanishi using a two-dimensional integral representation,
in terms of  a smooth                                                           
weight function $g$,  which carries the detailed dynamical information. 
A similar, but one-dimensional,  integral representation can be obtained for the Light-Front wave function in terms of the same weight function $g$.
By using the generalized Stieltjes transform, we first obtain $g$ in terms of the Light-Front wave function in the complex plane of its arguments. 
Next, a new integral equation for the Nakanishi weight function $g$ is derived for a bound state case. 
It has the standard form $g= N \, g$, where ${N}$ is a two-dimensional integral operator. 
We give the prescription for obtaining the kernel $ N$  starting with the kernel $K$ of the Bethe-Salpeter equation.  
The derivation is valid for any  kernel given by an irreducible Feynman amplitude.
\end{abstract}
\vspace{-0.5cm}
\end{frontmatter}

\section{Introduction}\label{Intr}

The Bethe-Salpeter (BS) equation \cite{SB_PR84_51} for a few-body system takes  into account the relativity and, implicitly, 
all the possible intermediate states.
 As demonstrated in \cite{bs-long} for the case of the ladder kernel -- which displays the strongest (pole) singularities --   it is possible to find  the bound  states and the scattering solutions in Minkowski space by a direct numerical calculations, 
 in spite of the many singularities appearing in the integral and inhomogeneous terms. 

There exist however  another efficient approach to solve this equation, proposed in \cite{KusPRD95} and developed in a series papers \cite{bs1,bs2,FrePRD14,FSVPRD,dPaPRD16,GutPLB16}, 
which relies on the Nakanishi representation \cite{nakanishi} of the BS amplitude   $\Phi(k,p)$ in terms of a non-singular  weight function $g(\gamma,z)$ as follows:
\begin{equation}\label{eq1}
\Phi(k,p) =  \int_{-1}^{1} dz^{'} \int_{0}^{\infty} d\gamma{'}  \frac{g(\gamma^{'}, z^{'})}{\left( \gamma^{'} + \kappa^2  -{k}^2 -p\cd {k}\,z^{'}-i\epsilon \right)^3},\,\,\,
\end{equation}
where 
$\kappa^2=m^2-\frac{M^2}{4},$
$m$ is the constituent mass and $M$ is the total mass of the bound state ($p^2=M^2$).
Once $g$ is known,  one can compute the BS  amplitude by means of (\ref{eq1})
and calculate other observables, like  the electromagnetic form factors (see  \cite{CarEPJA09}).  

The weight function $g$ determines also, via another integral representation, the light-front (LF) wave function  \cite{bs1}:
\begin{equation} \label{lf1}
\psi_{LF}(\gamma,z)={1-z^2\over 4}\;  \int_0^{\infty}  \frac{ g(\gamma{'}, z) d\gamma{'} }{ \left[ \gamma{'} + \gamma +  z^2 m^2 + \left( 1 - z^2 \right) \kappa^2 \right]^2 },
\end{equation}
In this one-dimensional representation, variable $z$  plays the role of a parameter, varying in the limits $-1\leq z \leq 1$.
The relations between variables ($\gamma,z$)  and the standard LF variables $(k_{\perp},x)$  are given by  $\gamma=k_{\perp}^2$ and $z=2x-1$.
By Eq. (\ref{lf1}) one can obtain the LF wave function and calculate its contribution to the form factors as well as the momentum distributions.  

In a series of  recent papers  \cite{bs1,bs2,FrePRD14,FSVPRD,nak,nak1}, two independent methods for finding $g$ have been  developed. 
In the  first one \cite{bs1,bs2,FrePRD14,FSVPRD}  the Nakanishi weight function $g$ is determined by solving the double integral 
equation,  previously established in \cite{bs1}:
\begin{small}
\begin{equation} \label{bsnew}
\int_0^{\infty}\frac{g(\gamma',z)d\gamma'}{\left[\gamma+\gamma' +z^2 m^2+(1-z^2)\kappa^2\right]^2} =
\int_0^{\infty}d\gamma'\int_{-1}^{1}dz'\;V(\gamma,z;\gamma',z') g(\gamma',z'),
\end{equation}
\end{small}
where the kernel $V$ is expressed via the BS kernel. 
The solution  $g$ of equation (\ref{bsnew}) was found  in the case of an OBE  ladder BS kernel  in \cite{bs1} and for the ladder plus cross-ladder one  in \cite{bs2,ff_cross}. 

Equation (\ref{bsnew}) contains integral terms in both sides, though the integral in the l.h. side is one-dimensional.
This fact generates some instability of  its numerical  solution, specially taking into account that when discretizing the left-hand side it results into  an ill-conditioned matrix \cite{bs1}.
An equation for  $g$,  getting rid of the left-hand side integral term of (\ref{bsnew}), was first derived in  \cite{FrePRD14}. 
This derivation is based on the uniqueness of the Nakanishi representation and it was fulfilled for the ladder kernel only. 
To clarify this point, let us represent  equation (\ref{bsnew}) in a symbolic form as: 
\begin{equation}\label{Lg_Vg}                     
 \hat{L}\; g=\hat{V} \; g\
 \end{equation}
In ref. \cite{FrePRD14},  the equation (\ref{Lg_Vg}) was first transformed into an equivalent one having the form \mbox{$L\cd (g-{\mathcal V}\cd g)=0$}.
This transformation used essentially  the properties of the ladder kernel $\hat{V}$.
Its straightforward application to the same equation with a non-ladder kernel would meet some  algebraic difficulties. 
The uniqueness theorem \cite{nakanishi} claims that if \mbox{$L\cd f=0$}, it follows that \mbox{$f=0$}. 
In this way, one obtains the equation 
\begin{equation}\label{g_Vg}
 g={ \hat{\mathcal V}} \; g\
 \end{equation}
written in terms of a unique integral operator ${\mathcal V}$. 
The main result of  \cite{FrePRD14} consisted in finding  equation  (\ref{g_Vg}) and the corresponding kernel ${\mathcal V}$. 
This equation was also numerically solved in \cite{FrePRD14} and the binding energies  were in very good agreement with the solutions of Eq. (\ref{bsnew}) found in \cite{bs1}. In addition, the stability of results relative to method and rank of discretization was much better.

In the second method  \cite{nak,nak1}, the Nakanishi weight function $g$ is extracted  from the 
"Euclidean"\footnote{The term "Euclidean" can be misleading and should be clarified. We mean the "Wick rotated" BS amplitude depending on imaginary $k_0$ component of the relative momentum, namely, on $k_0=ik_4$ with real $k_4$.} BS amplitude 
$\Phi_E$.
Since, as mentioned, the equation for  $\Phi_E$  can be solved rather easily and, in addition, 
$\Phi_E$ is accessible by the lattice calculations in a  full QFT framework, one can try to invert the representation (\ref{eq1}) 
and find in this way,  the Nakanishi weight function $g$.  
Once known, it can be used to calculate  the Minkowski  amplitude $\Phi_M$ and the observables in Minkowski space.
Notice that the Nakanishi weight function $g$ is the same for both the Minkowski and Euclidean BS amplitudes but depends 
on the total mass of the system $M$ through the parameter $\kappa$.
All the scalar products, resulting in the singularities, are contained in the denominator of (\ref{eq1}). 

Alternatively, 
since the LF wave function $\psi_{LF}$ can be also found relatively easy from the corresponding LF equation (given e.g. in \cite{cdkm}), 
one can try to find $g$ by inverting the representation (\ref{lf1}). 
Here again, the procedure is complicated by the instabilities  of the discretized equations (\ref{eq1}) and (\ref{lf1}) relative to $g(\gamma,z)$ 
since both of them belong to the class of so called ill-conditioned problems. 
We would like to  emphasize however that both equations are invertible in the mathematical sense, 
and so  they accept a unique solution $g$. Only its numerical inversion, when required with some accuracy, is far from being straightforward
 and requires some inspired methods.

The aim of this paper is to derive from (\ref{bsnew}) an equivalent equation written in the normal  form  $\mbox{$g=N\cd g$}$. 
To this aim, we will not rely on the uniqueness of the Nakanishi representation neither on the ladder kernel, 
but we will find the explicit form of the inverse operator $L^{-1}$ in  (\ref{Lg_Vg}) in a general case (Section \ref{Analytical_Inversion}).
We will first apply  this inverse  operation to (\ref{lf1})  and express  $g$ in terms of the LF wave function (Section \ref{sec3}).
Applying the same result to  equation (\ref{Lg_Vg}), we will derive in Section \ref{deriv} an explicit  representation  of $N=L^{-1}V$.
in terms of a one dimension integral on $V$.
Our derivation is valid for any kernel $K$ in the original BS equation or for any kernel $V$ in  equation (\ref{bsnew}).
In the case of a ladder kernel we will check the identity $N\equiv{\mathcal V}$.

\section{Analytical inversion of the Stieltjes kernel}\label{Analytical_Inversion}

\begin{figure}[h!]
\begin{center}
\epsfig{figure=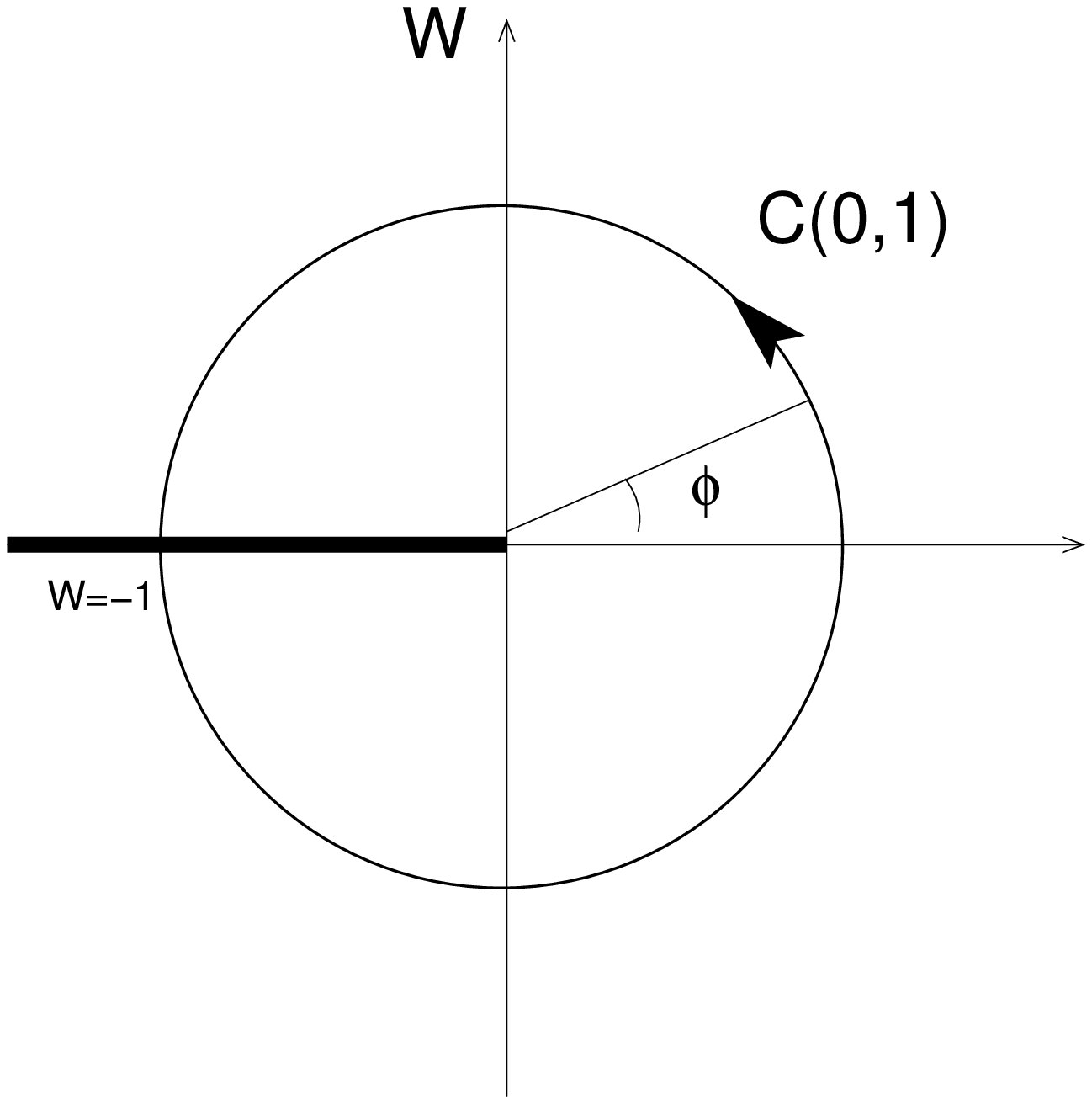,width=5cm} \hspace{2.cm}
\epsfig{figure=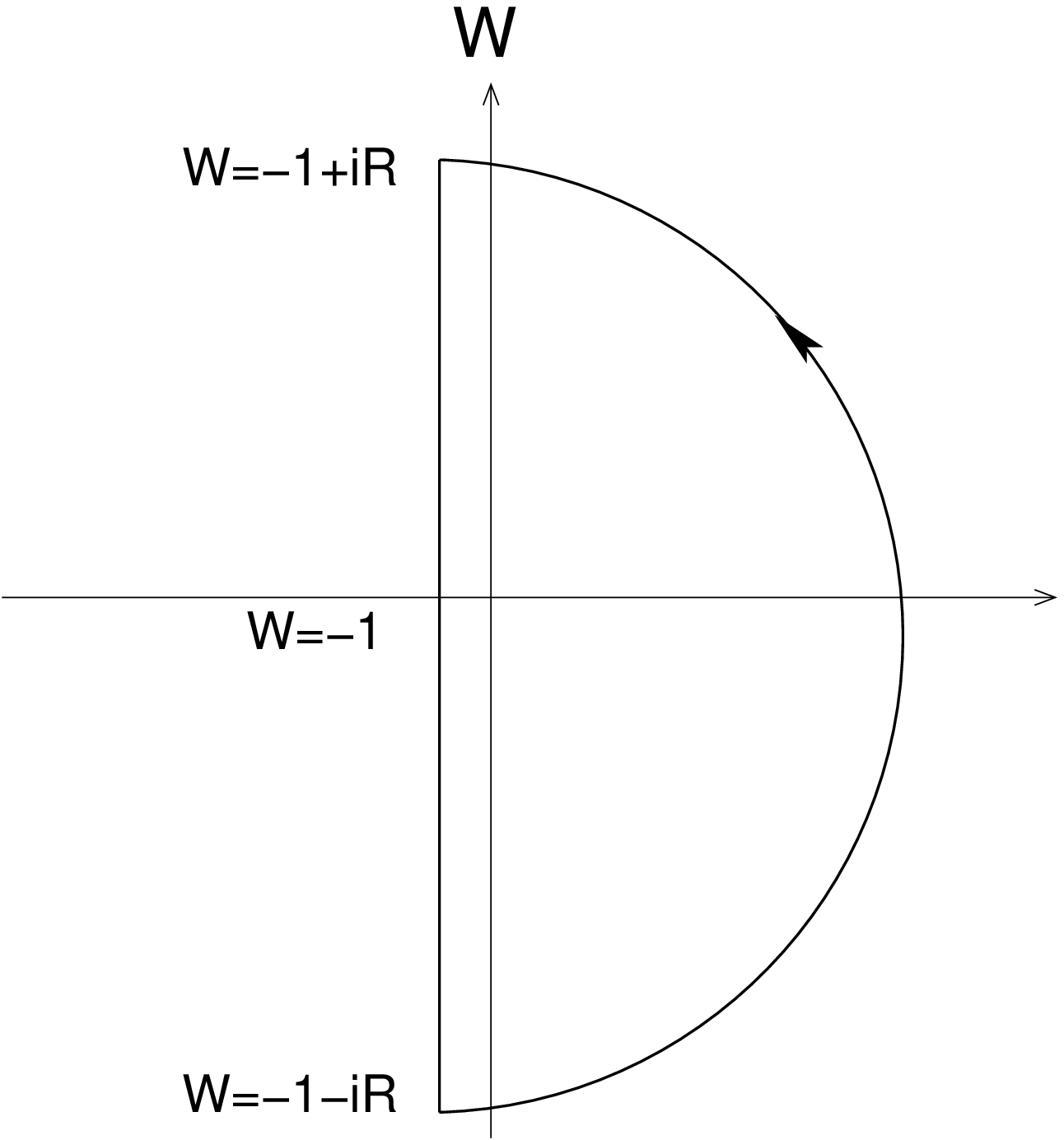,width=5cm}
\end{center}
\caption{Possible integration contours for  inverting the Sieltjes transform in Eq. (\ref{st2}).}\label{Contour_Stieljes}
\end{figure}

The results that follow are based on the observation, that  the left-hand-side of  (\ref{bsnew}), and  the right-hand-side of (\ref{lf1}),   can be trivially related to the Stieltjes transform:\footnote{We have reversed the notation $F\leftrightarrow G$ used in \cite{schw} in order to represent below $G$ by the Nakanishi weight  function $g$.}
\begin{equation}\label{st1}
F(x)=\int_0^{\infty}\frac{G(y)}{(y+x)^{\rho}} \; dy
\end{equation}
This integral transform can be analytically inverted \cite{Sumner_49,schw} for an arbitrary $\rho$. 
In the particular case $\rho=2$ its inversion is given by:
\begin{equation}\label{st2}
G(y)=\frac{y}{2\pi i}\int_{\mathcal C}F(yw)dw,
\end{equation}
where ${\mathcal C}$ is a contour in the complex plane of $w$ that starts and ends at the point $w=-1$ enclosing the origin in the counter-clockwise sense.
Two possible choices are illustrated in Fig. \ref{Contour_Stieljes}. 
We will take hereafter (left panel)  the unit circle centered at the origin $C(0,1)$ although the contour displayed in the right panel could be well adapted for further calculations.
The variable $w$ can then be parametrized as $w=\exp(i\phi)$ and the integration over the contour is reduced to the integration over $\phi$ in the limits $-\pi < \phi < \pi$.
That is
\begin{equation}\label{st3}
G(y)=\frac{y}{2\pi }\int_{-\pi}^{+\pi} d\phi \; e^{i\phi} \; F(y e^{i\phi} ) 
\end{equation}

Let us now consider an integral relation of the form
\begin{equation} \label{L}
f(\gamma)\equiv  \int_0^{\infty} \;d\gamma' \;L(\gamma,\gamma') g(\gamma')   = \int_0^{\infty} \;d\gamma' \; \frac{g(\gamma')}{ (\gamma'+\gamma+b)^2}  
\end{equation}
denoted symbolically as  $f = \hat{L} \;  g$.

By introducing  the shifted variable $\gamma''=\gamma+b$, a straightforward  application of  the inverse Stieltjes transform  (\ref{st3}) applied to  (\ref{L}) gives
\begin{equation} \label{L3}
g(\gamma)= \hat{L}^{-1} f=\frac{\gamma}{2\pi}\int_{-\pi}^{\pi}d\phi\, \;e^{i\phi}\; f(\gamma \; e^{i\phi}-b ).
\end{equation}
Though the proof of Eq. (\ref{st2}) is given  \cite{schw} in the general case (\ref{st1}), we will give  below a simple demonstration for the particular case
of  the operator   (\ref{L}) we are interested in.

To this aim, let us write (\ref{L3}) as 
\[ g= \hat{L}^{-1} \; f= \hat{L}^{-1} \; \hat{L} \; g  = \hat{U} \; g \equiv \int_0^{\infty} d\gamma' U(\gamma,\gamma';\epsilon) g(\gamma')  \]
kernel of the operator $\hat{U}\equiv  \hat{L}^{-1} \; \hat{L}$ 
is written for convenience in the form:
\begin{equation} \label{L4}
U(\gamma,\gamma';\epsilon)=\frac{\gamma}{2\pi}\int_{-\pi+\epsilon}^{\pi-\epsilon} \;d\phi\;   \frac{ \exp(i\phi)}
{\Bigl[\gamma \exp(i\phi)+\gamma'\Bigr]^2}
\end{equation}
by introducing the parameter $\epsilon$, being understood that
once the integration (\ref{L4}) is performed, the limit $\epsilon\to 0$ should be taken.
The integral (\ref{L4}) can be  analytically computed and  gives:
\begin{equation}\label{J1}
U(\gamma,\gamma';\epsilon)=\frac{\gamma\sin(\epsilon)}{\pi[(\gamma'-\gamma\cos(\epsilon))^2+\gamma^2\sin^2(\epsilon)]}.
\end{equation}
In the limit  $\epsilon\to 0$,  the kernel $U(\gamma,\gamma';\epsilon)$    is a sharp function  of $\gamma'$
with a maximum at $\gamma'=\gamma$ given by 
\[ U(\gamma,\gamma;\epsilon)\sim {1\over \epsilon}  \to \infty \]
Furthermore the integral:
\[ \int_0^{\infty}U(\gamma,\gamma';\epsilon)d\gamma'=1-\frac{\epsilon}{\pi}\to 1. \]
Therefore,  $U(\gamma,\gamma';\epsilon\to 0)$ is a representative of a  delta-function:
\begin{equation} \label{L5}
U(\gamma,\gamma';\epsilon\to 0)=\delta(\gamma'-\gamma)
\end{equation}
that is $\hat{U}$ is the unit operator 
$ \hat{U}=  \hat{L}^{-1} \; \hat{L} = 1 $
and so  (\ref{L3}) is the inverse of  (\ref{L}).

It is interesting to illustrate the inverse  equation (\ref{L3}) by  means of an anaytically solvable model.
This is provided, for simplicity in the case $b=1$, by the following pair $(f,g)$ satisfying Eq. (\ref{L}):
\begin{equation}
\label{eq4}
f(\gamma)=\frac{1}{\gamma^3}\left[\frac{\gamma(2+\gamma)}{(1+\gamma)}-2\log(1+\gamma)\right]
\end{equation}
\begin{equation}
\label{eq3}
g(\gamma)=\frac{1}{(1+\gamma)^2}
\end{equation}
By inserting $f$ given by (\ref{eq4})  into  Eq. (\ref{L3}), the integral can be performed analitically and we indeed reproduce $g$ from (\ref{eq3}). 
Care must be taken however when using  a numerical method to perform an integration which involves 
complex logarithms (see the remark at the end of next section).

\section{Computing the Nakanishi   weight function in terms of the light-front wave function}\label{sec3}

A first straigthforward application of  this result   can be done by inverting  Eq.   (\ref{lf1}), which express the LF
wave function $\psi_{LF}$ in terms of the Nakanishi weight function $g$.
By setting   in (\ref{L3}) 
\[   f={4\psi_{LF}\over 1-z^2}   \qquad  b=  m^2z^2 + (1-z^2)\kappa^2 \]
and introducing explicitly the $z$-dependence (which plays the role of a parameter)  on $g$, one has
\begin{equation}\label{g_Psi}
g(\gamma,z)  =  {\gamma\over 2\pi} {4\over 1-z^2} \; \int_{-\pi}^{\pi}   d\varphi \; e^{i\varphi}  \; \psi_{LF}[ \gamma e^{i\varphi} - m^2z^2 - (1-z^2)\kappa^2,z  ]
\end{equation}

The Nakanishi weight function is thus formally expressed in terms
of the LF wave function provided it is known in the complex plane of its first argument.

By inserting $g$ computed by   (\ref{g_Psi}) in Eq.  (\ref{eq1}) one can obtain the BS amplitude $\Phi$,  both in Euclidean ($\Phi_E$) and in Minkowski ($\Phi_M$) space.
Notice that the possibility to obtain the valence LF wave function from the Minkowski BS amplitude  was already established  some years ago by a projection technique 
\cite{cdkm,Tobias_BS_LF_PROJ}. 
The contents of Eq. (\ref{g_Psi})  can be considered as a way to realize the reciprocal procedure.

This result is remarkable and highly non trivial since it provides a link betwen two relativistic frameworks -- Light Front Dynamics  and Bethe-Salpeter equation -- which
have a  very different dynamical contents.
The key stone of this link is provided by the Nakanishi weight function $g$.

The interest of such a possibility in different fields of hadronic physics 
was emphasized in our previous works  \cite{nak,nak1}
where we presented an attempt to determine $g$ by  numerically inverting a discretized version of  Eqs. (\ref{eq1}) or (\ref{lf1}).
The corresponding linear systems turned to be ill-conditioned and, apart from  some analytical models like the ones described
in previous section -- Eqs. (\ref{eq4}) and (\ref{eq3}) --  its determination is not free from uncertainties.
Equation (\ref{g_Psi}) constitutes a formal prove of such an inversion.
 
The practical interest of (\ref{g_Psi}) deserves however  further investigations.
It relies on the knowledge of the LF wave function on the complex plane.
This is indeed technically accessible, as it has been done in \cite{Maris} for the Euclidean BS equation, but it is not an easy task and 
it is not clear that could be more efficient that the numerical inversions described in \cite{nak,nak1}. 
It is worth noticing that $g$ is real for bound states, a property which is far form being explicit in (\ref{g_Psi}).
This suggests that other integration contours than $C(0,1)$ in (\ref{st2})  taking benefit of this property, like the one on the right panel of Fig. 1 with $R\to\infty$, could be more suitable.

\section{Deriving the integral equation for $g$}\label{deriv}

By applying the 
integral transform (\ref{L3}) to both sides of the equation (\ref{bsnew}) \textcolor{red}{and using eq. (\ref{L5})}, we obtain
the following equation for the Nakanishi weight function $g$  in the canonical form
\begin{equation}\label{gNg}
g= \hat{L}^{-1}\hat{V} g\equiv \hat{N} g
\end{equation}
More explicitly
\begin{equation} \label{L6}
g(\gamma,z)=\int_0^{\infty}d\gamma'\int_{-1}^{1}dz'\;N(\gamma,z;\gamma',z') g(\gamma',z'),
\end{equation}
where 
\begin{equation} \label{L7}
N(\gamma,z;\gamma',z') =\frac{\gamma}{2\pi}\int_{-\pi}^{\pi} \;d\phi\;  e^{i\phi} \; V\Bigl(\gamma e^{i\phi}-z^2 m^2-(1-z^2)\kappa^2,z;\gamma',z'\Bigr).
\end{equation}

The operator $\hat{N}=\hat{L}^{-1}\hat{V}$ in (\ref{gNg}) exists if the corresponding integral (\ref{L7})  converges. 
Some divergence could eventually appear if the integration contour crosses a singularity of $V$ in the complex plane. 
In this case, the infinitesimal imaginary ''$-i\epsilon$'' term added to masses in the Feynman amplitudes, as usual,  is equivalent to a contour deformation  allowing to avoid crossing singularities. 
Since $V$ is related to a Feynman amplitude, it contains $-i\epsilon$, the kernel $N(\gamma,z;\gamma',z')$ is finite and the operator $\hat{N}=\hat{L}^{-1}\hat{V}$ in (\ref{gNg}) exists.

The equation (\ref{L6}) determines also a relation between the coupling constant $\alpha$ in the kernel $N$ and the total mass $M$ (or the binding energy $B=2m-M$) of the system or its spectrum (if any). As usual, the solution $g$ is found for a given $M$, and so although no explicitly written $g$ depends on the binding energy.

The relation between the original kernel $K$  appearing in the BS equation and the  kernel $V$ in (\ref{L7}) and (\ref{bsnew}) was derived in Ref. \cite{bs1}.
In  the case of a one boson exchange  kernel
\begin{equation} \label{K}
K(k,k')= -{g^2 \over (k-k')^2-\mu^2+i \epsilon },   \qquad g^2= 16\pi m^2 \alpha
\end{equation}
$V$ takes the form:
\begin{equation} \label{Kn}
V(\gamma,z;\gamma',z')=\left\{
\begin{array}{ll}
W(\gamma,z;\gamma',z'),&\mbox{if $-1\le z'\le z\le 1$}\\
W(\gamma,-z;\gamma',-  z'),&\mbox{if $-1\le z\le z'\le 1$}
\end{array}\right.
\end{equation}
where:
\begin{equation}\label{V1}
W(\gamma,z;\gamma',z')= + \frac{\alpha m^2(1-z)^2} {2\pi [\gamma +z^2 m^2+(1-z^2)\kappa^2 ]} \int_0^1 dv\; \left(\frac{v}{D}\right)^2
\end{equation}
and
\begin{eqnarray*}
D&=&v(1-v)(1-z')\gamma+v(1-z)\gamma'+v(1-z)(1-z')\Bigl[1+z(1-v) +v z'\Bigr]\kappa^2  \\
  &+&v\Bigl[(1-v)(1-z')z^2+v{z'}^2(1-z)\Bigr]m^2+(1-v)(1-z)\mu^2.
\end{eqnarray*}
The integral  (\ref{V1}) was calculated analytically in \cite{bs1} and reads:
\begin{eqnarray}\label{W}
W(\gamma,z;\gamma',z') &=&
\frac{\alpha m^2}{2\pi}  \frac{(1-z)^2}{\gamma+z^2m^2+(1-z^2)\kappa^2} \\
&\times& \frac{1}{b_2^2(b_+ -b_-)^3}  \left[ \frac{(b_+ -b_-)(2b_+ b_- -b_+ -b_-)}{(1-b_+)(1-b_-)}  +  2b_+ b_- \log \frac{b_+ (1-b_-)}{b_- (1-b_+)}\right] \nonumber
\end{eqnarray}
where
\begin{eqnarray}\label{bb}
b_0 &=& (1-z)\mu^2, \,\,\,\,\, b_\pm = -\frac{1}{2b_2} \;\left( b_1 \pm \sqrt{b_1^2-4b_0b_2}\right), \\
b_1 &=& \gamma+\gamma' - (1-z)\mu^2 - \gamma' z -\gamma z' +(1-z')\left[z^2m^2+(1-z^2)\kappa^2\right], 
\nonumber\\
b_2 &=& -\gamma (1-z')-(z-z') \left[  (1-z)(1-z')\kappa^2+(z+z'-zz') m^2 \right].
\nonumber
\end{eqnarray}
By inserting  (\ref{Kn})  in the integral (\ref{L7}), the kernel $N$ of  Eq. (\ref{L6}) will be determined.
Notice however that the benefit of using the analytic expression (\ref{W}) with respect to the numerical integration (\ref{V1}) 
is not always  granted, neither
from the computational point of view nor from the ambiguities arising when using multivalued complex integrands.

As mentioned in the Introduction,  the canonical form of Eq. (\ref{bsnew}) was first derived   in \cite{FrePRD14} 
 under the hypothesis of the ladder kernel.
These authors  wrote this equation --  Eq. (24) of  \cite{FrePRD14}  --  in the symbolical form (\ref{g_Vg})
with a kernel ${\mathcal V}$ that,  when written in a form close to (\ref{Kn}),  reads
\footnote{Following to \cite{Gianni}, we have corrected a misprint in  the kernel sign in \cite{FrePRD14} and modify accordingly
 Eqs. (\ref{Vf}) and (\ref{hp}).}
\begin{equation}\label{Vf}
{\mathcal V}(\gamma,z;\gamma',z') = + \frac{\alpha m^2}{2\pi}\times
\left\{
\begin{array}{ll}
\displaystyle{h(\gamma,-z;\gamma',-z')}, &\quad \mbox{if $-1\le z'\le z\le 1$} \\
\displaystyle{h(\gamma,z;\gamma',z')}, &\quad  \mbox{if $-1\le z\le z'\le 1$}
\end{array}
\right.
\end{equation}
Function $h$,  obtained from $h'$ given in  \cite{FrePRD14} by Eqs. (27) and (28), takes the form: 
\begin{equation}\label{hp}
h(\gamma,z;\gamma',z')=  \theta(\eta) \; P( \gamma,z,\gamma',z' )   +  Q(\gamma',z' )
\end{equation}
where
\[ P(\gamma,z,\gamma',z' )  =   \frac{B}{\gamma A\Delta}\frac{ 1+z }{( 1+z' )}  -  C(\gamma,z,\gamma',z' )   \]
with
\begin{eqnarray}
&&A (\gamma',z' )                         =   \frac{1}{4}{z'}^2 M^2+\kappa^2+\gamma', \,\,\,\,\,
B(\gamma,z,\gamma',z' )         = \mu^2+\gamma'-\gamma\frac{1+z'}{1+z},   \cr
&& C(\gamma,z,\gamma',z' )        =  \int_{y_-}^{y_+}\chi(y)dy,
\,\,\,\,
Q(\gamma',z' )                         =  \int_{0}^{\infty}\chi(y)dy ,\cr 
&&\Delta (\gamma,z,\gamma',z' ) = \sqrt{B^2-4\mu^2A}, 
\end{eqnarray}
The integrand on $C$ and $Q$ reads
\begin{eqnarray}
\chi(y) = \frac{y^2}{[y^2+A+y(\mu^2+\gamma')+\mu^2]^2}   \label{chi}
\end{eqnarray}
and the integration limits are given by
$y_{\pm} = \frac{-B\pm\Delta}{2A}.$
The argument $\eta$ of the $\theta$-function in the first term of (\ref{hp}) is:
\begin{equation}\label{arg}
\eta=-B-2\mu\sqrt{A}=\gamma\frac{1+z'}{1+z}-\mu^2-\gamma'-2\mu\sqrt{\frac{1}{4}{z'}^2 M^2+\kappa^2+\gamma'}.
\end{equation}

In order that  Eqs. (\ref{gNg}) and (\ref{g_Vg}) could represent the very same equation, the kernels $N$ and ${\mathcal V}$ must be identical to each other.
However the complexity of their analytical forms, specially taking into account that  $N$ results  from the integral  (\ref{L7}), 
makes any direct comparison beyond the scope of the present paper. 
 
 To prove this identity we must entirely rely on numerical calculations.
 More precisely, we will substitute the kernel $V$ defined by  (\ref{Kn}) in Eq. (\ref{L7}), 
 calculate the integral over $\phi$ for a selected set of external variables ($\gamma,z,\gamma',z'$), determine  $N(\gamma,z,\gamma',z')$  and compare the result with 
 ${\mathcal V}( \gamma,z,\gamma',z')$ given by (\ref{Vf}). 
 
 We have considered two of these sets, corresponding  to negative and positive values of $\eta$,  the argument of the $\theta$ function  appearing in Eq. (\ref{hp}). 
 The kernel OBE parameters were fixed to $\alpha=1,\; m=1,\;\mu=0.15,\;M=1.9$.
 
 For the first set of variables
\begin{equation}\label{set1}
\gamma=0.1,\;\gamma'=1,\;z=0.2,\;z'=0.35,
\end{equation}
one has $\eta=-1.24<0$, and the $\theta$ function in  (\ref{hp}) does not contribute to the result. 
The numerical calculation gives: 
\begin{eqnarray}\label{Vt1}
N&=&0.1153980591510\\
{\mathcal V}&=&0.1153980591509
\nonumber
\end{eqnarray}
As one can see that the $N$ and ${\mathcal V}$ kernels coincide with each other within 12 digits, the computer accuracy. 

For the second set of variables
\begin{equation}\label{set2}
\gamma=0.5,\;\gamma'=0.1,\;z=0.2,\;z'=0.35,
\end{equation}
$\eta=0.27>0$, and $\theta$ function in  (\ref{hp})  contributes. The numerical calculations give: 
\begin{eqnarray}\label{Vt2}
N         &=&        -0.05 31 89 08 58 160 \\
{\mathcal V}&=&        -0.05 31 89 08 58 158
\nonumber
\end{eqnarray}
Here again, the kernels  coincide with each other within the same accuracy. 
Note that these calculations were performed using the analytical expression for  the $W$ function (\ref{W}).

The following warning concerns  numerical calculations involving complex multivalued functions. 
When using  {\it Mathematica}  or fortran packages to evaluate the integral  (\ref{L7}) with  analytical expressions like Eq. (\ref{W}) in the integrand, 
one should take care with the implicit  cuts   on the multivalued functions (e.g. $\log(z)$, $\sqrt{z}$, etc.)  implied in these tools.
For example, the result  depends on whether or not we apply to r.h. side of (\ref{W}) the {\it Mathematica} command \mbox{PowerExpand}. 
The kernel (\ref{Vt2}) for  set (\ref{set2}) was obtained after applying \mbox{PowerExpand}; without applying it, the two kernels are different.\footnote{\footnotesize 
\begin{eqnarray*}
\mbox{Examples:}
\phantom{\int_0^1 dx \frac{z(1-2z)}{z+(1-2z)x} = 
z\,\mbox{Log}\left[\frac{1-z}{z}\right]\phantom{]} }
&&
\\
\int_0^1 dx \frac{z(1-2z)}{z+(1-2z)x} = 
z\,\mbox{Log}\left[\frac{1-z}{z}\right]\phantom{]} /. \{z \to 0.1 + i0.2\}&=&\phantom{-}0.407 + i0.151
\\
\neq\mbox{PowerExpand}\left[z\,\mbox{Log}\left[-\frac{1}{z}(z - 1)\right]\right] /. \{z \to 0.1 + i0.2\} &=&-0.850+ i0.780 
\end{eqnarray*}
The result depends on the order of factors in the argument of $\log$, on using or not the \mbox{PowerExpand}  (!) command  and on how $\log$ is defined. 
Chose e.g.  $z=\exp\left(-i\frac{\pi}{4}\right)=0.707(1-i)$. 
The value of $\log(z)$ depends on the branch one takes: either $\log(z)=\log\left[\exp\left(-i\frac{\pi}{4}\right)\right]=-\frac{i\pi}{4}=-0.785i$, 
or $\log(z)=\log\left[\exp\left(i\left(-\frac{\pi}{4}+2\pi\right)\right)\right]=\frac{7i\pi}{4}=5.498i$. The same happens with $\sqrt{z}$ in (\ref{bb})  when using Eq. (\ref{W}). 
This reflects the fact that we are dealing with multivalued function. 
Since the numerical calculation of an integral, like  (\ref{V1}),  is unambiguous we take it as the correct value.}
Whereas the result for set (\ref{set1}) is independent on using \mbox{PowerExpand} or not.
Note that the equation (\ref{V1}) for $W$ in the integral form does not contain this ambiguity. After numerical calculation of the double integral (over $\phi$ and $v$), $N$ coincides with ${\mathcal V}$
given by (\ref{Vf}), though with less precision.

The results described in this Section  confirm  the identity of the two kernels $N$ and ${\mathcal V}$, despite their very different formulation and analytical expressions.
This identity  between these two kernels, proved in the case of a ladder approximations, suggests  to use the integral form (\ref{L7}) over $V$
as an alternative way to compute ${\mathcal V}$ in a more involved dynamical context or for scattering case.
Indeed, the validity of Eq. (\ref{L6}) does not rely on the uniqueness ansatz and  has been established 
under the same conditions than Eq. (\ref{bsnew}), that is for any kernel $K$ inserted in the BS equation.

The same remark concerning the real character of $g$ discussed at the end of  the previous Section holds here in the definition of  $N$ itself through
the complex integral  (\ref{L7}). The solution $g$ as well as the kernel are, in the bound state case, both real
and it would be interesting to take benefit on that and make it explicit with an appropriate choice of the contour in Fig. 1.

\section{Conclusion}\label{concl}

We have considered the Nakanishi integral representation of the  Bethe

weight function $g$.
These integral transforms are difficult to invert since they result into ill-conditioned linear systems. 

By using an analytic inversion of the Stieltjes integral transform we have
first obtained a closed form expressing the Nakanishi weight function in terms of the Light-Front wave function in the complex plane.

Starting with a Bethe-Salpeter equation for the weight function $g$ in the form $Lg=Vg$ -- see Eq. (\ref{bsnew}) -- previously established in \cite{bs1,bs2}
we have used the same inversion procedure to write it in the canonical form $g=Ng$ -- see Eq. (\ref{L6}) ---  with
the kernel $N$ given by an integral of kernel $V$.

This equation  generalizes, to the case of a  kernel given by any set of  irreducible Feynman graphs, the equation found in \cite{FrePRD14} 
valid only for the ladder kernel and using the uniqueness anzatz. 
This generalization strongly expands the applicability of the Nakanishi representation to find the solution of the Bethe-Salpeter equation, 
as well as its applications.

It is worth noticing that the validity of applying the inverse Stieltjes transform for deriving Eq. (18) and  solving in this way the Bethe-Salpeter
equation relies on the fact that  the Bethe-Salpeter kernel is generated by irreducible Feynman diagrams. 
This allows to use the Nakanishi representation for the Bethe-Salpeter amplitude  and obtain in this way  the equation (\ref{bsnew}). 
When using an arbitrary kernel -- e.g. constructed by hand or inspired on phenomenological grounds, however providing converged results --
the Nakanishi representation for the corresponding amplitude could not be defined and therefore  the applicability of the Stieltjes transform  would be not guaranteed.

The first evident application    of the method developed here,  could be to find the solution $g$ for the ladder plus cross-ladder kernel, previously obtained  from Eq. ({\ref{bsnew}) in  \cite{bs2,ff_cross}.
The equation (\ref{L6}) can also be extended straightforwardly to the two-fermion system starting, for example, from \cite{dPaPRD16}.

The possibility to apply an analytic inversion of the Stieltjes transform can be useful in other fields
of nuclear and hadronic physics, where the use of  integral transforms was  pioneered 
by  \cite{Efros_SJNP_1985,Efros:2007nq,Giussepina_EFB23_2016},
in order to avoid the instabilities of the numerical inversion.

\section*{Aknowledgements}
We are indebted to  G.~Salm\`e  for useful discussions.  TF thanks Conselho Nacional de Desenvolvimento Cient\'ifico e Tecnol\'ogico(CNPq),  
Coordena\c c\~ao de Aperfei\c coamento de Pessoal de N\'ivel Superior 
(CAPES) and Funda\c c\~ao de Amparo \`a Pesquisa do Estado de S\~ao Paulo (FAPESP) of Brazil.
V.A.K. thanks the support  of Funda\c c\~ao de Amparo \`a Pesquisa do Estado de S\~ao Paulo (FAPESP),  the grant \#2015/22701-6.
He is also sincerely grateful to group of theoretical nuclear physics of ITA, S\~{a}o Jos\'e dos Campos, Brazil, for kind hospitality during his visit.

\end{document}